\documentstyle[11pt,newpasp,twoside]{article}
\def\lcdm{$\Lambda$CDM}
\def\gsim {\lower .1ex\hbox{\rlap{\raise .6ex\hbox{\hskip .3ex
        {\ifmmode{\scriptscriptstyle >}\else
                {$\scriptscriptstyle >$}\fi}}}
        \kern -.4ex{\ifmmode{\scriptscriptstyle \sim}\else
                {$\scriptscriptstyle\sim$}\fi}}}
\def\lsim {\lower .1ex\hbox{\rlap{\raise .6ex\hbox{\hskip .3ex
        {\ifmmode{\scriptscriptstyle <}\else
                {$\scriptscriptstyle <$}\fi}}}
        \kern -.4ex{\ifmmode{\scriptscriptstyle \sim}\else
                {$\scriptscriptstyle\sim$}\fi}}}
\def\kms{\ {\rm km\,s^{-1}}}

\def\hMsun{h^{-1}M_{\odot}{\ }}
\def\Mvir{M_{\rm vir}}
\def\mvir{M_{\rm vir}}

\def\cvir{c_{\rm vir}}

\def\rs{r_{\rm s}}
\def\Rvir{R_{\rm vir}}

\def\vmax{V_{\rm max}}
\def\Vmax{V_{\rm max}}
\def\vmax{V_{\rm max}}

\def\edcomment#1{\iffalse\marginpar{\raggedright\sl#1\/}\else\relax\fi}
\reversemarginpar

\begin{document}
\title{Evolution of Mass Distribution in CDM Halos}
\author{Joel R. Primack}
\affil{Physics Department, University of California, Santa Cruz, CA
95060 USA}

\begin{abstract}
On the basis of a new convergence study of high-resolution N-body
simulations, my colleagues and I now agree that the Navarro, Frenk, \&
White (1996) density profile $\rho_{NFW}(r) \propto r^{-1}
(r+r_s)^{-2}$ is a good representation of typical dark matter halos of
galactic mass.  Comparing simulations of the same halo with numbers of
particles ranging from $\sim10^3$ to $\sim10^6$, we have also shown
that $r_s$, the radius where the log-slope is -2, can be determined
accurately for halos with as few as $\sim10^3$ particles.  Based on a
study of thousands of halos at many redshifts in an Adaptive
Refinement Tree (ART) simulation of a cosmological volume in a
$\Lambda$CDM cosmology, we have found that the concentration $\cvir
\equiv \Rvir/\rs$ has a log-normal distribution, with $1\sigma$
$\Delta (\log \cvir) = 0.18$ at a given mass, corresponding to a
scatter in maximum rotation velocities of $\Delta \Vmax/\Vmax = 0.12$.
The average concentration declines with redshift at fixed mass
as $\cvir(z) \propto (1+z)^{-1}$.  This may have important
implications for galaxy rotation curves.  Finally, we have found that
the velocity function determined from galaxy luminosity functions plus
luminosity-velocity relations agrees with the predictions from our
$\Lambda$CDM simulations.  But we also note that the very limited
evolution with redshift of the velocity function predicted by
$\Lambda$CDM conflicts with the data that is becoming available on the
number density of bright galaxies unless there is significant
evolution of the luminosity-velocity relation at $z>1$.
\end{abstract}

\section{Introduction}

In this talk, I review some of the recent work by my collaborators
(especially my former PhD student James Bullock) and me on the
distribution of dark matter in galaxy-size halos and its evolution
with redshift.  I summarize results from several recent papers, in
particular Bullock et al. (1999), Gonzalez et al. (2000), Sigad et
al. (2000), Klypin et al. (2000), and Bullock et al. (2000).  Avishai
Dekel in his talk summarized related work by our group on the
distribution of angular momentum in dark matter halos.

\section{New results on the centers of dark matter halos}

The ART code (Kravtsov, Klypin, \& Khokhlov 1997) starts with a
uniform grid treated with a Particle-Mesh algorithm, but refines all
high-density regions using an automated refinement mechanism, with the
time-step correspondingly reduced.  Extensive new tests of the ART
code and comparison with other simulation codes are presented in
Kravtsov (1999) and Knebe et al. (2000).  In an earlier paper,
Kravtsov et al. (1998) (discussed also in Primack et al. 1999), we
analyzed ART simulations which resolved dozens of halos in small
volumes for CDM, CHDM, and $\Lambda$CDM cosmologies.  We concluded
that the central density behavior is $\rho \propto r^{-\gamma}$, with
$\gamma$ typically $\sim0.3$ but ranging from about 0 to 1 for
different halos.  There we used results from ART simulations with a
maximum formal dynamic range of $256\times2^6=16,384$, corresponding
to a best formal resolution (size of the smallest refinement mesh
cell) of $l_{\rm mesh} \sim0.5 \, h^{-1}$ kpc, and we used only
results for $\geq 2 l_{\rm mesh}$.  This was because the convergence
study that we described in that paper showed that {\it for a fixed
mass resolution} the halo density profiles converged at $2 l_{\rm
mesh}$ as we increased the force resolution.

For reliable results at the centers of dark matter halos it is also
necessary to consider the effects of mass resolution.  We have now
done a more careful analysis of the convergence by simulating the {\it
same} galaxy-mass halo with increasing numbers of particles.  Our
highest resolution runs achieved a formal spatial dynamical range of
$2^{17}=131,072$; the simulation was run with 500,000 steps at the
highest level of refinement.  Here we concentrated on the $\Lambda$CDM
cosmology.  We used a new version of the ART code with particles of
various masses, so that we could put the lowest-mass particles in the
region of the box containing the halo we were interested in and higher
mass particles farther away.  Our results show that there is no change
in the simulated density profile as we increase the number of
particles by a very large factor, down to a radius of at least 4 times
the formal force resolution and containing at least 200 simulation
particles.  We find that $\rho_{NFW}(r)$ is a good fit to our highest
resolution halos, although we show that several other popular analytic
formulas also give good fits.  In particular, the NFW formula is a
good (better than 10\%) fit to our halos down to 0.01 of the virial
radius, which corresponds to $\gsim 1 h^{-1}$ kpc for the dwarf and
LSB galaxies that are often compared to model predictions.  It is
hardly possible to measure rotation curves of such galaxies at 
smaller radii, and even if one could do so there are various physical
effects that would make it difficult to interpret the results in terms
of a density profile.  Note that, although the logslope of the NFW
density profile is -1 in the limit as $r \longrightarrow 0$, at 0.01
of the virial radius the logslope is considerably steeper.  For
simulated halos the actual density profiles have features that deviate
from smooth fitting formulas such as NFW, and at least some of these
features appear to reflect the merging history of the halos.

Based on our new convergence study, we no longer trust the results
reported in Kravtsov et al. (1998) concerning the very centers of
halos.  In particular, our results concerning the shallow central
slopes depended on trusting our simulations between 2 and 4 times the
formal resolution.  However, all the results in Kravtsov et al. (1998)
at radii greater than 4 times the formal resolution should still be
valid, including the scatter in profile shapes and the agreement
between the $V_{max}$ vs. $ r_{max}$ relations of simulated dark halos
and those of dark-matter-dominated dwarf and LSB galaxies.  However,
since the HI data on some of these galaxies was affected by
beam-smearing (van den Bosch et al. 2000) and H$\alpha$ data is now
available for some of them (Swaters et al. 2000, Swaters \& van den
Bosch 2000; cf. contributions by Swaters and van den Bosch to these
proceedings), it would be worthwhile to repeat this analysis.

The new H$\alpha$ data resolves much of the concern (e.g., Flores \&
Primack 1994) with rotation curves of dark-matter-dominated galaxies
contradicting the cuspy halos from CDM simulations.  The main related
concern about too many small satellite halos compared to the number of
observed satellite galaxies in the local group has been most
convincingly addressed by the model of Bullock, Kravtsov, \& Weinberg
(2000), presented in two posters at this meeting.  They show that only
those small halos that have collapsed before the epoch of reionization
will accrete gas and subsequently be able to produce stars, and that
the numbers of the resulting small satellites --- both those observed,
and those subsequently accreted by the Milky Way --- are in excellent
agreement with observations.

\section{Distribution and evolution of halo concentration}

\begin{figure}
\vskip 5.5cm
{\includegraphics{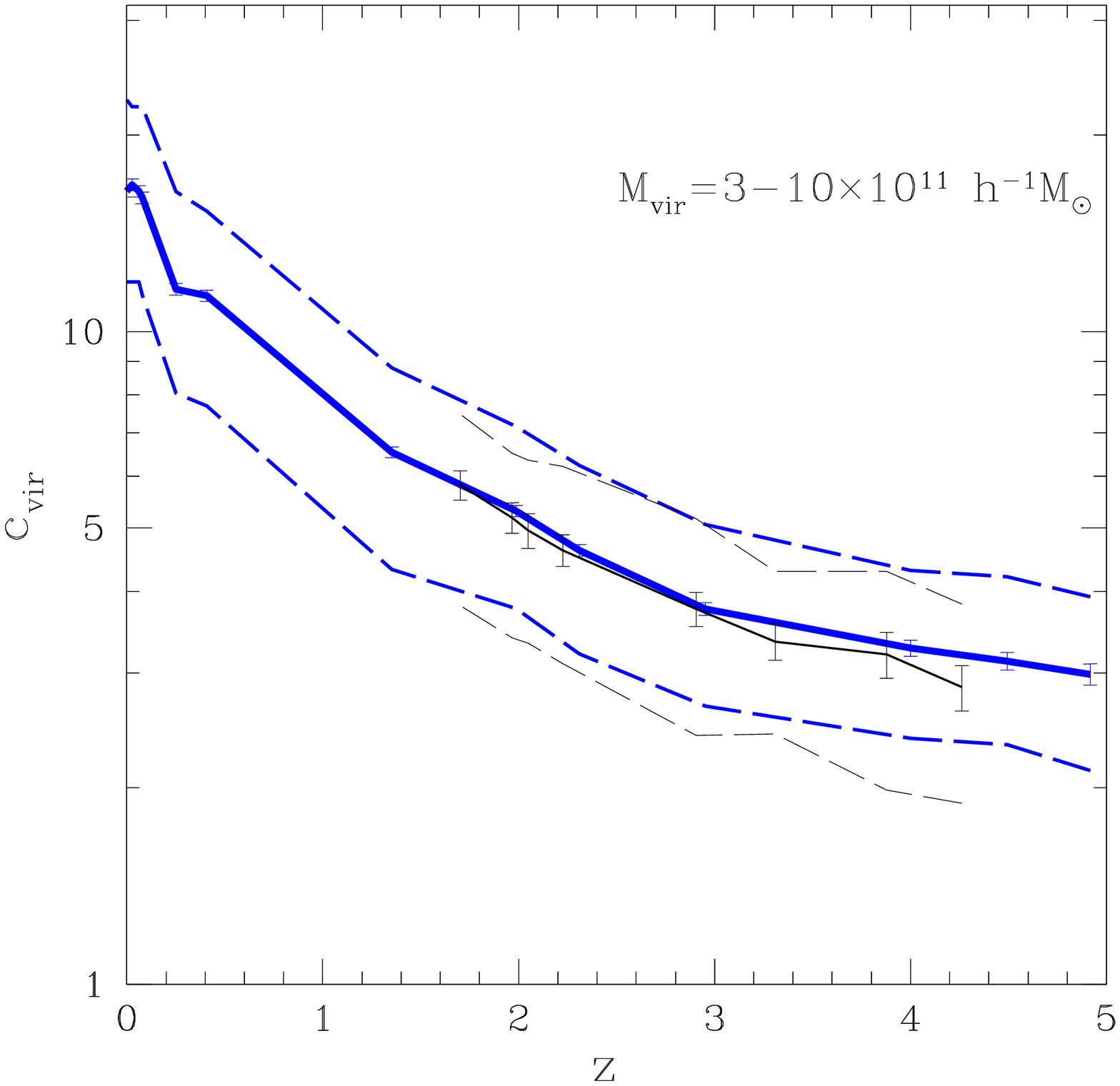}}
{\includegraphics{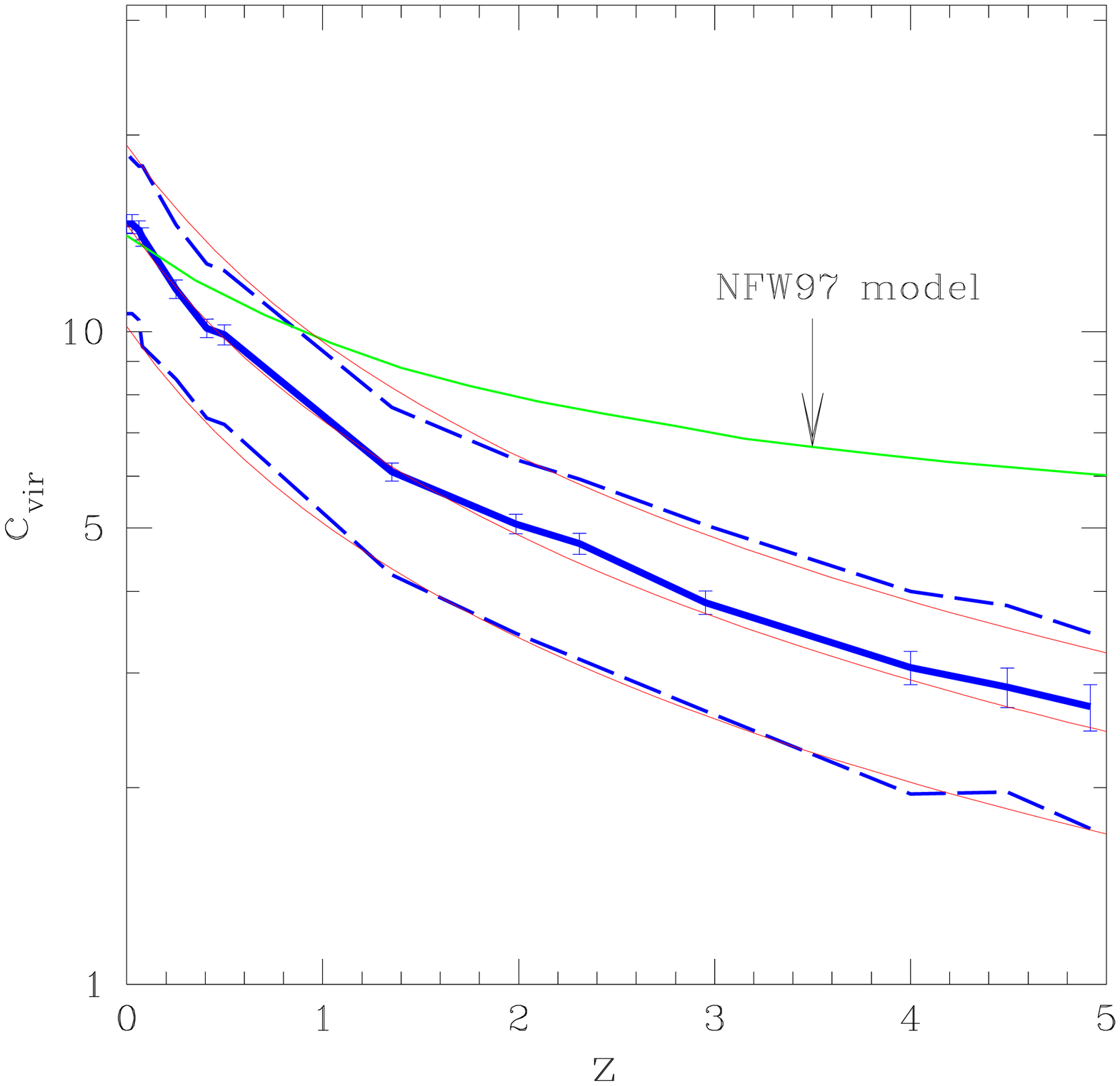}}
\caption{(a) Convergence test for $\cvir$ evolution and scatter.
Shown is a comparison of $\Mvir = 3-10 \times 10^{11} \hMsun$ haloes
simulated using our main simulation (thick lines) and a second
simulation with $8$ times the mass resolution (thin lines).  The solid
lines and errors reflect the median and Poisson uncertainty
respectively.  The dashed lines reflect the estimated intrinsic
scatter.  There is no evidence for significant deviations in either
the measured median or scatter as the mass resolution is increased.
(b) Concentration as a function of redshift for distinct halos of a
fixed mass, $\mvir=0.5-1.0\times10^{12}\hMsun$.  The median (heavy
solid line) and intrinsic $68\%$ spread (dashed line) are shown.  The
behavior predicted by the NFW97 model is marked.  Our revised 
model (available at www.astronomy.ohio-state.edu/~james/CVIR/parts.html)
for the median and spread for $8 \times 10^{11} \hMsun$ halos
(thin solid lines) reproduces the results from the simulations rather 
well.}
\label{fig:profiles}
\end{figure}

The convergence study of Klypin et al. (2000) shows that ART
simulation halos with as few as $\sim10^3$ particles can be used to
get accurate values for the NFW parameter $\rs$ (where the logslope of
$\rho(r)$ is -2), as long as the NFW fit is restricted to begin at a
sufficiently large radius.  In Bullock et al. (1999), thousands of
halos from a 60 $h^{-1}$ Mpc volume simulation of $\Lambda$CDM were
analyzed at many redshifts to determine the distribution of halo
concentration as a function of redshift.  As Fig. 1 (a) shows, a
simulation with the same number of particles ($256^3$) in a 30
$h^{-1}$ Mpc box which was run to redshift $z=1.7$ gave essentially
identical results for the distribution of halo concentrations for
halos of $(3-10)\times 10^{11} h^{-1} M_\odot$ at overlapping
redshifts, despite the fact that identical mass halos had 8 times as
many particles in the smaller box simulation.

We define the halo concentration as $\cvir \equiv \Rvir/\rs$, where
the virial radius $\Rvir$ is defined as the radius within which the
mean density is the virial overdensity $\Delta_{\rm vir}$ times the
average density at that redshift $\rho_{\rm ave}$.  The value of
$\Delta_{\rm vir}$, determined from the spherical top hat collapse
approximation, depends on the cosmology and the redshift (see Bullock
et al. 1999 for details).  For the $\Omega_m=0.3$ $\Lambda$CDM
cosmology that we discuss, $\Delta_{\rm vir} \approx 340$ at $z=0$.
(The definition above differs from that of NFW, who defined $c \equiv
R_{200}/\rs$, where $R_{200}$ corresponds to $\Delta=200$, appropriate
for an Einstein-de Sitter cosmology.)

The dark matter halos in a fixed mass range at any redshift have
approximately a log-normal distribution of concentrations (Jing 2000,
Bullock et al. 1999).  In Bullock et al. (1999) we present a simple
analytic model for the concentration of dark matter halos.  Like the
model presented in Navarro, Frenk, \& White (1997), it relates the
concentration of a halo to the epoch when a certain fraction of the
final mass in that halo had already collapsed; and like the NFW model,
it correctly predicts the average halo concentration as a function of
halo mass at redshift $z=0$.  But unlike the NFW model, it also
correctly predicts the concentration as a function of redshift (see
Fig. 1 (b)).  In particular, it gives the $\cvir \propto (1+z)^{-1}$
behavior that is evident in Fig. 1.

Our model also correctly accounts for the observed 1-$\sigma$ spread
of concentrations (shown in Fig. 1 by the upper and lower dashed
curves) in terms of the spread in halo formation epochs due to the
Gaussian distribution of fluctuation amplitudes in CDM.  The spread in
halo concentrations has a large effect on galaxy rotation curve
shapes, comparable to the effect of the well-known log-normal
distribution of halo spin parameters $\lambda$.  Frank van den Bosch
(2000) showed, based on a semi-analytic model for galaxy formation
including supernova feedback, that the spread in $\lambda$ mainly
results in movement along the Tully-Fisher line, while the spread in
concentration results in dispersion perpendicular to the TF relation.
Remarkably, he found that the dispersion in CDM halo concentrations
produces a TF dispersion that is consistent with the observed one.

In order to compare theoretical rotation curves to those of actual
galaxies, it is necessary to take into account the effects on the dark
matter halo of the dissipative collapse of the baryons that form the
disk.  In the papers (Blumenthal et al. 1986, Flores et al. 1993) that
first discussed the effect of baryonic infall on galaxy rotation
curves, we considered $z=0$ galaxy disks to have formed at $z\sim1$.
Since in our study of CDM halo evolution we find that the
concentration evolves $\propto (1+z)^{-1}$, this would clearly result
in lower concentration than if we used the $z=0$ halo properties.
This is a topic that requires further investigation before we can
properly compare concentrations of observed galaxies with predictions
of CDM models.

\section{Galaxy velocity function from luminosity function and
luminosity-velocity relations}

A strength of CDM models is that it is possible to calculate the
number density of halos with given properties.  Although it is also
possible to predict the number density of galaxies as a function of
their luminosity by means of semi-analytic models (e.g., Somerville \&
Primack 1999), calculating luminosities of galaxies in CDM halos
requires treatment of the poorly understood processes of gas cooling,
star formation, and feedback, and many simplifications are necessary.
The velocity function is a much simpler connection between theory and
observation.  Luminosity-velocity relations derived from observations
--- the Tully-Fisher and Faber-Jackson relations --- allow one to
construct approximate galaxy velocity functions from the observed
galaxy luminosity functions.  These are approximate since it is
necessary to average over galaxy inclination for spiral galaxies, and
for surveys in which the morphologies of galaxies were not determined
it is necessary to make the approximation that all galaxies are spiral
galaxies.  However, in Gonzalez et al. (2000) we showed that the
resulting uncertainty in the number density of galaxies with rotation
velocity $\sim 200 \kms$ is only about a factor of 2.  We found that
the observational number density determined this way agrees well with
that predicted in \lcdm\ with $\Omega_m=0.3$ if we don't take into
account the effect of baryonic infall, although it is perhaps a bit
low when we do take this into account.  New surveys, in particular the
Sloan Digital Sky Survey, will determine the luminosity function much
more accurately, and it will be important to measure the corresponding
Tully-Fisher relation and compare the resulting velocity function to
the predictions of cosmological models.

The fact that the luminosity-velocity relations work so well in the
nearby universe raises the question whether they will continue to hold
at higher redshift.  In a recent paper (Bullock et al. 2000, based in
part on the detailed analysis of \lcdm\ velocity functions in Sigad et
al. 2000), we point out that in the observationally favored \lcdm\
cosmology with $\Omega_m=0.3$ and $\sigma_8=1$, the number density of
halos with maximum circular velocity $\vmax=200\kms$ increases only by
about 30\% between redshift 0 and 3, and then declines back to the
local value by $z=5$.\footnote{There is a more dramatic increase with
redshift in the number density of halos with virial velocity of
$200\kms$, but the decrease in concentration with increasing redshift
compensates for this and results in much less increase in the number
density at fixed $\vmax$.  The number density of halos with
$\vmax=200\kms$ also increases out to $z\sim2$ in open or Einstein-de
Sitter CDM.  Note that halos with fixed $\vmax$ have decreasing mass
at higher redshift.}  If the luminosity-velocity relations continue to
hold at these higher redshifts, the implication is that the comoving
number density of bright galaxies should increase.  This appears to
contradict the data from the Northern Hubble Deep Field (Dickinson
2000), which shows a dramatic decrease in the number of bright
galaxies above redshift $z\sim1.4$.  So this suggests that the
luminosity-velocity relations may change or perhaps even become
stochastic at higher redshifts.  The Keck DEEP survey will attempt to
measure the internal velocities of $\sim60,000$ galaxies at $0.7\lsim
z \lsim 1.5$, which should allow a direct test of this.  Understanding
the evolution of the luminosity-velocity relations may perhaps clarify
their physical origin.

{\vskip 0.2cm}

\noindent {\bf Acknowledgments}.  I thank all my collaborators for
many enlightening discussions on these topics.  I am grateful for
support from NASA and NSF grants at UCSC, and from a Humboldt Award at
the Max Planck Institute for Physics, Munich.


\end{document}